\documentclass[a4paper]{jpconf}
\usepackage{graphicx}
\usepackage{slashed}
\usepackage{amsmath, amsthm, amssymb}
\begin{document}
\title{Photon emission and decay from generic Lorentz invariance violation}

\author{H. Mart\'inez-Huerta and A. P\'erez-Lorenzana}

\ead{hmartinez@fis.cinvestav.mx and aplorenz@fis.cinvestav.mx}

\address{Physics Department, Centro de Investigacion y de Estudios Avanzados del I.P.N., \\ Apartado Postal 14-740, 07000, Mexico City, Mexico. }

\begin{abstract}

One of the most studied approaches in phenomenology to introduce the breaking of Lorentz symmetry is the generic approach. This consist on the modification of the free particle  dispersion relation by the addition of an extra power law term of order $n$ on energy or momentum. Using this approach in the photon sector, we have calculated the generic rates for vacuum Cherenkov radiation and photon decay, for any order $n$,  at leading order. Explicit results for the decay and emission rates for the lowest values of $n$ are also presented.

\end{abstract}

\section{Introduction}

Lorentz invariance violation (LIV), mainly motivated by quantum gravity and string theories  \cite{QG1,QG2, QG3, QG4}, represents an interesting sector in the search for physics beyond the Standard Model. Since Lorentz symmetry has a fundamental role in the construction of the model, the derived physics form LIV extensions tends to be unique and energy dependent in order to preserves the standard physics unchanged \cite{DIS2, DIS1}. 
Additionally, LIV evidence is usually expected to increase with the energy and distance.

Currently, the most energetic known phenomena are cosmic rays (CR). They can reach energies of several decades of EeV \cite{XMAX,GZK} and can travel astrophysical distances 
before detection. Moreover,  their study has shown a strong development in recent decades. Therefore, properly studied LIV consequences could be identified in the current CR observatories and experiments. 

Accordingly to the previous ideas, in this note we present  a phenomenological first order correction, using a generic approach, of the emission rates for two processes that could have a significant impact on cosmic particle propagation: photon decay and vacuum Cherenkov radiation.

The generic mechanism for the introduction of LIV, frequently used in the literature \cite{DIS2,DIS1, DIS3,DIS4, GUNTER-PH, VCR, GUNTER-PD, LIV-proc},
converges to an explicit not Lorentz invariant
term added to the free particle Lagrangian density that will generate a correction to the dispersion relation
\begin{equation}\label{eq_S}
    S_{a} = E_a^2 - p_a^2 = m_a^2 \pm \alpha_{a,n}A^{n+2},
\end{equation}
where $E_a$ and $p_a$ stand for the four momenta  associated with an $a$ particle species. $A$ can take the form of $E$ or $p$ for specific models, nevertheless, for  $m_a\ll\{E, p \} $, the ultrarelativistic limit, any particular choice of $A$ will be equivalent. The coefficient $\alpha_{a,n}$ in Eq.~(\ref{eq_S}), parametrizes the particle  dependent LIV correction, where $n$ expresses the correction order to the mass shell codified in the energy or momentum expression in $A$. 
The generic $\alpha_{n}$ is frequently inversely related to $n$-th power of $E_{QG}$, the scale of quantum gravity (QG) or the scale of the new physics beneath, which is expected to be close to  $10 ^ {19}$~GeV. Several methods are used in the search for LIV signals, some of them can lead to lower limits to $E_{QG}$ \cite{E3,E2,E1, HESS-LIV,FERMI-LIV,GRB-LIV, HAWC-LIV}.

In the next section,  we have applied the LIV generic correction in Eq.~(\ref{eq_S}) for photons to find photon emission and decay rate for the processes depicted in Figure 1 and 2, corrected at first $\alpha_{n}$ order but for any $n$. Such processes are forbidden in the standard theory by energy momentum conservation, but under the LIV hypothesis they can be permitted and can be used in the search for LIV signatures. 
In Figure 1, we show the emission of a single photon by a charged particle that propagates in vacuum, commonly named  vacuum Cherenkov radiation \cite{DIS2, GUNTER-PH,VCR} due its role as an energy loss mechanism in charged particle propagation. Figure 2 shows LIV photon decay \cite{DIS2, GUNTER-PH, GUNTER-PD},  
this process is kinematically allowed and motivated by the LIV extra term in Eq.~(1). 
The simplest process will produce an electron - positron pair. Once the general expression for the corrected rates had been presented, we shall used them to show their behaviour for particular values of $n$.

\begin{figure}[h]
	\begin{minipage}{14pc}
	    {\centering
		\includegraphics[width=9pc]{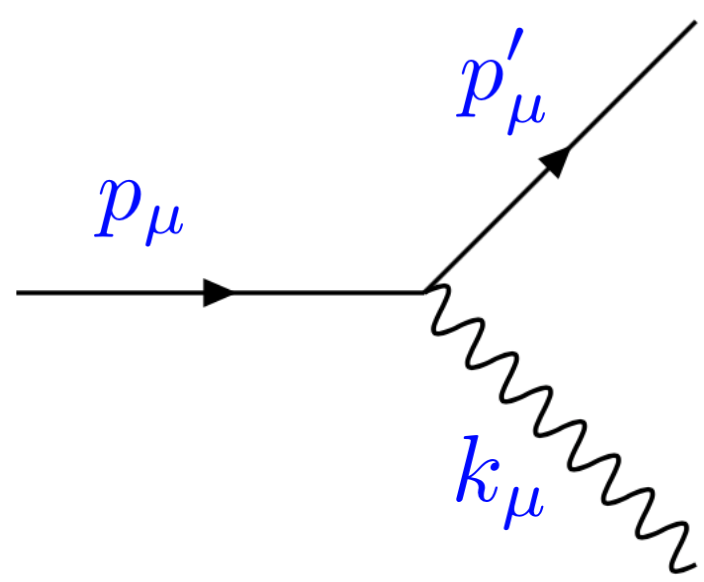}
		\caption{\label{fig1}Diagram for LIV vacuum Cherenkov radiation.}}
	\end{minipage}\hspace{8pc}%
	\begin{minipage}{14pc}
	    {\centering
		\includegraphics[width=9pc]{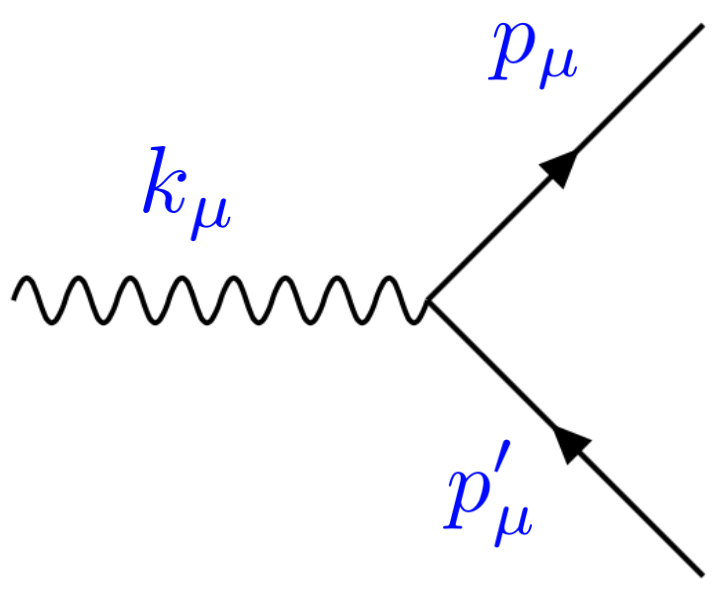}
		\caption{\label{fig2}Diagram for LIV photon decay.}}
	\end{minipage} 
\end{figure}

\section{LIV corrected rates}

Following the diagrams in Figures 1 and 2, photon four momenta will be named $(\omega,\mathbf{k})$, whereas prime and not prime notation  stand for the fermion  momenta, as indicated. A proton would be assumed for the first and an $e^+-e^-$ pair for the last.
They have been chosen due their relevance for cosmic~\cite{XMAX,MIXED2} and gamma rays~\cite{CMBAngelis}.
Hence, the corrected square amplitude at first LIV order in $\alpha_n$ will be given by:
\begin{equation}\label{vertex}
	\frac{1}{2}\sum_{spin} |M|^2 = e^2 |4m_a^2 - \alpha_{n} k^{n+2} | ,
\end{equation}
where orthogonality relation is assumed for photon polarizations at first approximation, and $k=|\mathbf{k}|$. The absolute value above is taking in order to ensure physical congruence due the limits and the generality in the sign of 
$\alpha_{n}$. For a more detailed calculation  see  Ref.~\cite{Proc2}.

Using Eq.~(\ref{vertex}) and the LIV corrected energy for photons, $\omega (k,\alpha_n) = k\sqrt{1+\alpha_{n}k^{n}}$,
we derived the  modified decay and emission rates on an arbitrary preferential frame for any order $n$.
For vacuum Cherenkov radiation, we found that 
\begin{equation}\label{eq_VCR_gamma}
	\Gamma_{a\rightarrow a\gamma}^{(n)} =   \dfrac{e^2}{4\pi} \dfrac{1}{4E_a} 
 	\int_0^\theta  \sum_{k_i}\frac{|4m_a^2 - \alpha_n k_i^{(n+2)}| }{\omega(k_i,\alpha_{n} )} 
	   \frac{k_i^2 \sin\theta d\theta}{| p_a \cos\theta -k_i  - \frac{(1+\frac{2+n}{2}\alpha_n k^n_i)}{\sqrt{1+\alpha_{n}k^n_i}} \sqrt{k_i^2 + E_a^2-2k_ip_a\cos\theta}|},
\end{equation}
where $k_i$ are the non zero photon momenta modes from the corrected energy-momenta conservation. For $n=0$ we have:
\begin{equation}\label{eq_k0}
	k_0^{(n=0)} = \frac{1}{p\cos\theta} \big[E^2(\alpha_0+1) - p^2\cos^2\theta\big],
\end{equation}
and for $n=1$,
\begin{equation}\label{eq_kplus}
	k_\pm^{(n=1)} =  \frac{1}{2p_a\cos\theta}
	\left( E_a^2   \pm \sqrt{E_a^4 +  \frac{4p_a\cos\theta}{\alpha_{1}}(E^2_a - p_a^2\cos^2\theta)}\right).
\end{equation}
Since $k_-$ is non physical, it will not  be  used  in our phenomenological approach.

Likewise, for LIV photon decay of order $n$ we get the rate
\begin{equation}\label{eq_PD_gamma}
	\begin{aligned}
	\Gamma_{\gamma\rightarrow e^+e^-}^{(n)} =   \dfrac{e^2}{4\pi}  \frac{|4m_e^2 - \alpha_{n}k^{n+2}| }{4 \omega(k,\alpha_{n})} 
	\int_0^{\pi} \sum_{p=p_\pm} \frac{p^2\sin\theta d\theta }{| (k\cos\theta - p)E_e - p\sqrt{k^2 + E_e^2-2kp\cos\theta}  | }, 
	\end{aligned}
\end{equation}
where the momenta modes from the corrected energy-momenta conservation and for any $n$ are
\begin{equation}\label{disc}
	p_{\pm}= \frac{1}{2(\alpha_nk^n + \sin^2\theta)}\left( \alpha_n k^{n+1}\cos\theta 
	 \pm \sqrt{\alpha_n^2 k^{2n+2}\cos^2\theta-4(\sin^2\theta+\alpha_n k^n) (1+\alpha_nk^n)m^2} \right) .
\end{equation}
Here, $e$ stands for the electron charge, $\theta$ for the angle between final particles and  $p_a=\sqrt{E_a^2-m_a^2}$ for the given particle $a$ absolute momentum.
Photon decay expression are  generic for any fermion pair in the final state, provided the corresponding mass is used, while the emission rate is, from an phenomenological approach, generic for any charged particle or nucleus with spin $= 1/2$.  
There exist vacuum Cherenkov radiation and photon decay rates obtained from different LIV approaches. For instance, expressions from the minimal Standard-Model extension with spontaneous breaking of Lorentz symmetry
\cite{SME} and from the introduction of Lorentz violating operators of dimensions four and six can be found in Refs.~\cite{TWO-side, CROSS}.

Integration on Eqs.~(\ref{eq_VCR_gamma}) and (\ref{eq_PD_gamma}), for different $n$, were numerically performed. The results are depicted in Figures 3 and 4 for Vacuum Cherenkov radiation and LIV photon decay respectively. A small angle approach is taken for vacuum Cherenkov radiation in order to clarify the threshold. Results for fixed $n$ and different $\alpha_{1}$ for both processes can be found in \cite{Proc2}.

\begin{figure}[h]
	\begin{minipage}{17pc}
	    {\centering
		\includegraphics[width=1.\textwidth]{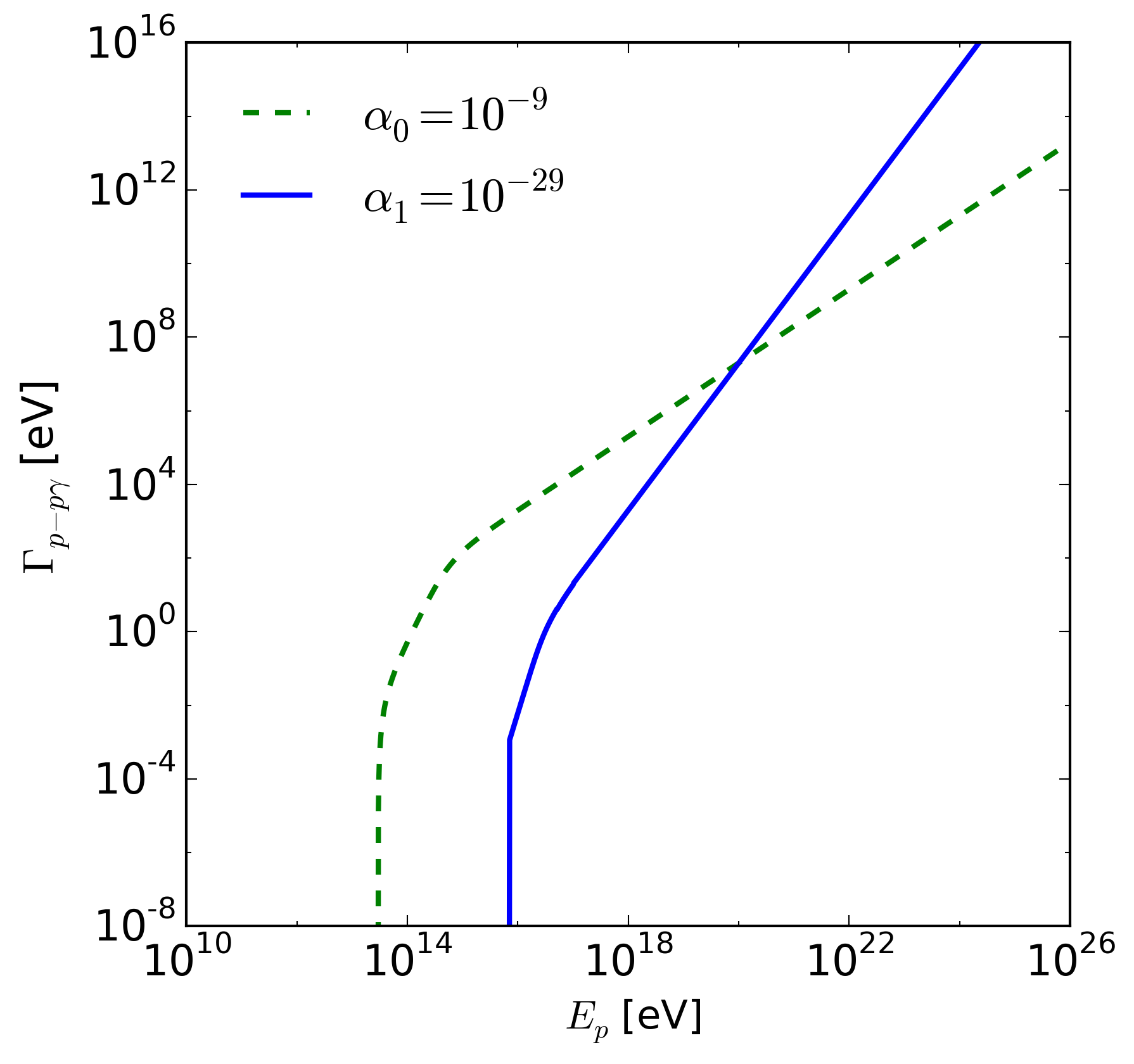}
		\caption{\label{fig3}Emission rate for LIV vacuum Cherenkov radiation for n=0,1, $m_{a} = m_{proton}$ and $\theta_{max}=10^{-2}$. 
		}}
	\end{minipage}\hspace{3pc}%
	\begin{minipage}{17pc}
	    {\centering
		\includegraphics[width=1.\textwidth]{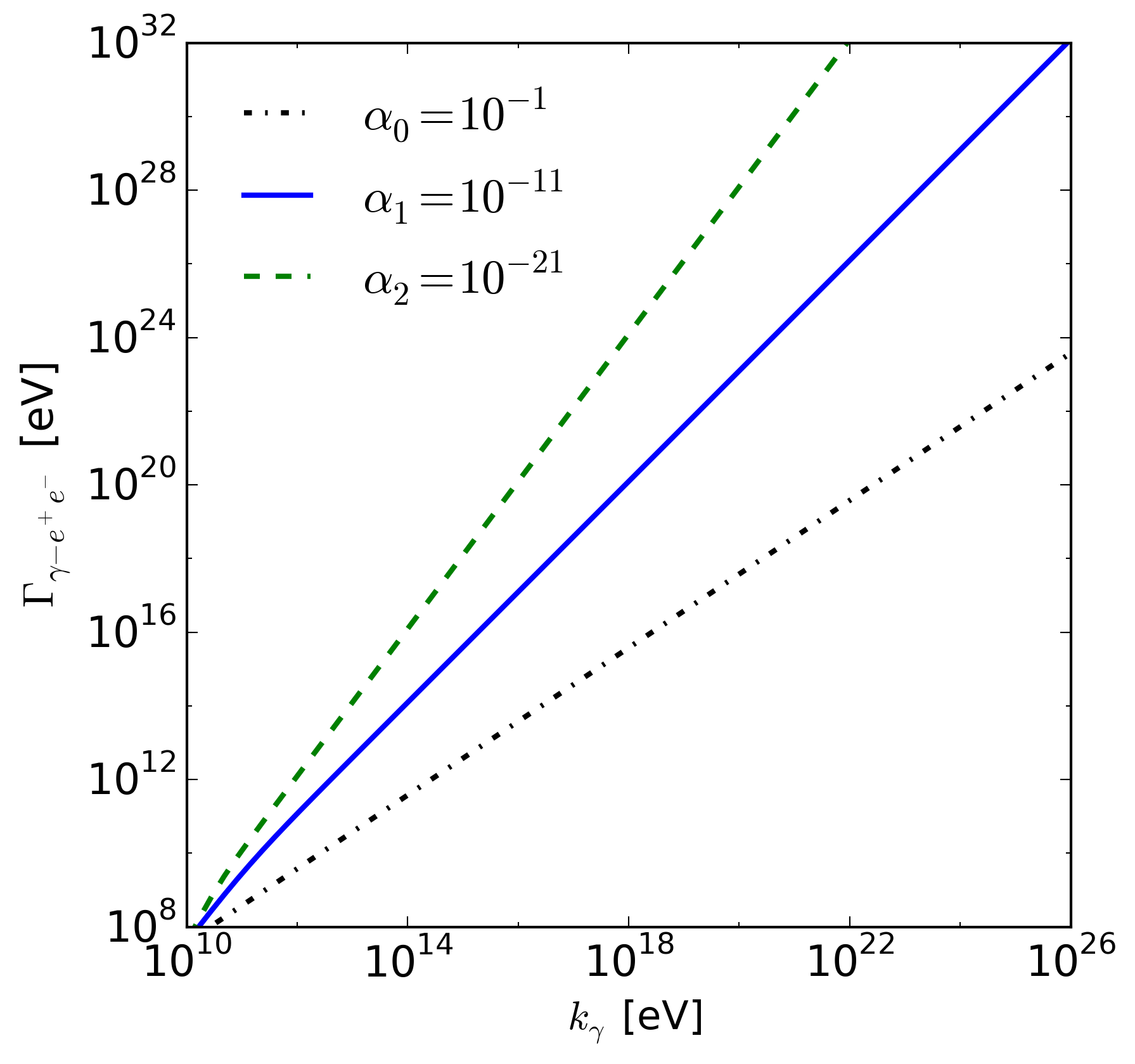}
		\caption{\label{fig4}Decay rate for LIV photon decay into electron positron pairs, for n=0,1,2. 
		}}
	\end{minipage} 
\end{figure}

As it can be seen in Figs. 3 and 4, the phenomena are sensitive to the LIV term. Both rates grow with the primary particle energy and they show a different slope with the order of the correction $n$. For  vacuum Cherenkov emission a phenomenological threshold is required to turn off the process and  preserve  the physics below it unaffected. We have proposed $E \ge (m^2/\alpha_0)^{1/2}$ for $n=0$ and $E \ge (4m^2/\alpha_1)^{1/3}$ for $n=1$. In LIV photon decay case, there is also a threshold that turns off the process at low energies but it naturally comes from energy-momentum conservation in the Lorentz conserving fermion sector and the corrected LIV photon.

\section{Conclusions}
We have found a generic first order LIV correction for every order $n$ to the emission and decay rates for vacuum Cherenkov radiation and LIV photon decay. Unlike Lorentz invariant theory, both processes are permitted under LIV hypothesis. As expected, their possible consequences  increases with the energy. They show the same phenomenological behaviour at the first orders in $n$ and can lead to different expected physics at the most energetic scenarios, such as cosmic rays, but we will present such an analysis in a future work.

\ack
This work was partially supported by Conacyt grant No. 237004. We are also grateful to Conacyt for the support trough the
Mesoamerican Center for Theoretically Physics.

\section*{References}

\bibliography{bibfile}

\providecommand{\newblock}{}
\begin{thebibliography}{10}
\expandafter\ifx\csname url\endcsname\relax
  \def\url#1{{\tt #1}}\fi
\expandafter\ifx\csname urlprefix\endcsname\relax\def\urlprefix{URL }\fi
\providecommand{\eprint}[2][]{\url{#2}}

\bibitem{QG1}
Ellis J, Mavromatos N~E and Nanopoulos D~V 2000 {\em Phys. Rev.\/} D {\bf 61}
  027503

\bibitem{QG2}
Gambini R and Pullin J 1999 {\em Phys. Rev.\/} D {\bf 59} 124021

\bibitem{QG3}
Ellis J, Mavromatos N and Nanopoulos D 2000 {\em Gen. Rel. Grav.\/} {\bf 32}
  127--144

\bibitem{QG4}
Ellis J, Mavromatos N, Nanopoulos D and Volkov G 2000 {\em Gen. Rel. Grav.\/}
  {\bf 32} 1777 -- 1798

\bibitem{DIS2}
Coleman S and Glashow S~L 1999 {\em Phys. Rev.\/} D {\bf 59} 116008

\bibitem{DIS1}
Amelino-Camelia G, Ellis J, Mavromatos N~E, Nanopoulos D~V and Sarkar S 1998
  {\em Nature\/} {\bf 393} 763--765

\bibitem{XMAX}
Aab A {\em et~al.\/} (Pierre Auger Collaboration) 2014 {\em Phys. Rev.\/} D
  {\bf 90} 122005

\bibitem{GZK}
Abraham J {\em et~al.\/} (Pierre Auger Collaboration) 2010 {\em Phys. Lett.\/}
  B {\bf 685} 239--246

\bibitem{DIS3}
Amelino-Camelia G 2001 {\em Nature\/} {\bf 410} 1065--1067

\bibitem{DIS4}
Ahluwalia D~V 1999 {\em Nature\/} {\bf 398} 199

\bibitem{GUNTER-PH}
Galaverni M and Sigl G 2008 {\em Phys. Rev. Lett.\/} {\bf 100} 021102

\bibitem{VCR}
Gagnon O and Moore G~D 2004 {\em Phys. Rev.\/} D {\bf 70} 065002

\bibitem{GUNTER-PD}
Galaverni M and Sigl G 2008 {\em Phys. Rev.\/} D {\bf 78} 063003

\bibitem{LIV-proc}
Mart\'inez H and P\'erez-Lorenzana A 2013 {\em Journal of Physics: Conference
  Series\/} {\bf 468} 012005

\bibitem{E3}
Ellis J~R, Mavromatos N~E, Nanopoulos D~V and Sakharov A~S 2003 {\em Astron.
  Astrophys.\/} {\bf 402} 409--424

\bibitem{E2}
Albert J {\em et~al.\/} (MAGIC, Other Contributors) 2008 {\em Phys. Lett.\/}
  {\bf B668} 253--257 (\textit{Preprint} \eprint{0708.2889})

\bibitem{E1}
Ellis J~R, Mavromatos N~E, Nanopoulos D~V, Sakharov A~S and Sarkisyan E~K~G
  2006 {\em Astropart. Phys.\/} {\bf 25} 402--411 [Erratum: Astropart.
  Phys.29,158(2008)] (\textit{Preprint} \eprint{0712.2781})

\bibitem{HESS-LIV}
Abramowski A {\em et~al.\/} (HESS Collaboration) 2011 {\em Astropart. Phys.\/}
  {\bf 34} 738–747

\bibitem{FERMI-LIV}
Vasileiou V, Jacholkowska A, Piron F, Bolmont J, Couturier C, Granot J, Stecker
  F~W, Cohen-Tanugi J and Longo F 2013 {\em Phys. Rev.\/} D {\bf 87} 122001

\bibitem{GRB-LIV}
Xu H and Ma B~Q 2016 {\em Phys. Lett.\/} B {\bf 760} 602--604

\bibitem{HAWC-LIV}
Nellen L (HAWC) 2015 {\em Proc. Sci., ICRC2015 (2016) 850\/} (\textit{Preprint}
  \eprint{1508.03930})

\bibitem{MIXED2}
Aab A {\em et~al.\/} (Pierre Auger Collaboration) 2014 {\em Phys. Rev.\/} D
  {\bf 90} 122006

\bibitem{CMBAngelis}
De~Angelis A, Galanti G and Roncadelli M 2013 {\em Mon. Not. R. Astron. Soc.\/}
  {\bf 432} 3245--3249

\bibitem{Proc2}
Mart\'nez-Huerta H and P\'erez-Lorenzana A 2016 {\em Journal of Physics:
  Conference Series\/} {\bf 761} 012035

\bibitem{SME}
Colladay D and Kostelecky V~A 1998 {\em Phys. Rev. D\/} {\bf 58} 116002

\bibitem{TWO-side}
Klinkhamer F~R and Schreck M 2008 {\em Phys. Rev. D\/} {\bf 78} 085026

\bibitem{CROSS}
Grigory~Rubtsov P~S and Sibiryakov S 2012 {\em Phys. Rev. D\/} {\bf 86} 085012

\end{thebibliography}

\end{document}